\documentclass[5p]{elsarticle}

\usepackage[utf8x]{inputenc}
\usepackage{units}
\usepackage[numbers]{natbib}
\usepackage[english]{babel}
\usepackage{blindtext}

\usepackage{tikz}
\usetikzlibrary{shapes,arrows}

\tikzstyle{block} = [rectangle, draw, 
    text width=7em, text centered, rounded corners, minimum height=3em]
\tikzstyle{blocktrans} = [rectangle,
  text width=6em, text centered, rounded corners, minimum height=4em]
\tikzstyle{blocksmall} = [rectangle, draw, 
    text width=6em, text centered, rounded corners, minimum height=2em]
\tikzstyle{decision} = [diamond, draw,
    text width=4.5em, text badly centered, node distance=3cm, inner sep=0pt]
\tikzstyle{line} = [draw, -latex']

\begin{document}
\bibliographystyle{elsarticle-num}

\begin{frontmatter}
\title{Recent integral cross section validation measurements at the ASP facility}

\author[UK]{L. W. Packer\corref{cor1}}
\ead{lee.packer@ccfe.ac.uk}
\author[AWE]{S. Hughes}
\author[UK]{M. Gilbert}
\author[UK]{S. Lilley}
\author[UK]{R. Pampin}

\address[UK]{Culham Centre for Fusion Energy, Culham Science Centre, Abingdon, Oxfordshire, OX14 3DB, UK}

\address[AWE]{AWE, Aldermaston, Reading, Berkshire, RG7 4PR, UK}

\cortext[cor1]{Corresponding author}

\begin{abstract}

This work presents new integral data measured at the ASP 14 MeV neutron irradiation facility at Aldermaston in the UK, which has recently become available for fusion-related work through the CCFE materials programme. Measurements of reaction products from activation experiments using elemental foils were carried out using gamma spectrometry in a high efficiency, high-purity germanium (HPGe) detector and associated digital signal processing hardware. Following irradiation and rapid extraction to the measurement cell, gamma emissions were acquired with both energy and time bins.

Integral cross section and half-life data have been derived from these measurements. Selected integral cross section values are presented from the measurement campaigns. Details of the data processing approach and outputs generated are highlighted for measurement of the \textsuperscript{186}W(n,2n)\textsuperscript{185m}W reaction---a selected short-lived reaction resulting from W foil irradiations; C/E results are reported along with the associated uncertainties and compared using the SAFEPAQ-II tool against existing available data.  

\end{abstract}


\end{frontmatter}

\section{Introduction}
When ranking the importance of technological challenges that need to be tackled in order to develop fusion power plants, the problem of determining the exact behaviour of materials through exposure to high intensity, high energy neutrons produced in such devices is certainly towards, if not at, the top of the list. The absence of available high energy neutron irradiation testing facilities with sufficiently high fluxes that are needed for material qualification means that, for now, progress in fusion technology is heavily reliant on simulations to predict materials performance. Among the range of material behaviours that need to be assessed are their nuclear performance during and after irradiation, which include associated activation and gas production quantities: activation analysis tools such as FISPACT and its supporting nuclear data, which are integral parts of the European Activation System \cite{Forrest2007_EASY2007,Forrest2009,Packer2011}, currently provide one of the best routes to do this.

To be able to give assurance in the results of activation calculations that are required for application to nuclear technologies, the supporting nuclear data libraries must be validated. The validation exercise that followed the production of EAF-2007 \cite{Forrest2007_EASY2007} is the most recent and most complete exercise of this type for an activation library \cite{EAFVAL}. It included a comparison of the evaluated differential cross section data with integral data for a given reaction measured over a range of neutron spectra. However, in the work, which was compiled using the SAFEPAQ-II tool \cite{Forrest2001,SAFE10}, only 470 reactions---out of a total of 65,565 reactions in EAF-2007---had both integral and differential data. A subset of these reactions, 217, were described as validated, meaning that the supporting differential and integral data are in agreement. A larger number of the remaining reactions had some limited supporting experimental data, whilst the vast majority of the reactions contained within the library were derived from nuclear model-based codes such as TALYS \cite{Koning2008}.

Experimental work carried out at the ASP 14 MeV irradiation facility \cite{Packer2012,AWEASP} has enabled CCFE to develop the capability to  contribute to the pool of validation experiments that are collected and assembled through the international EXFOR \cite{EXFOR} database maintained by the IAEA. Validation experiments are needed to support existing and successor nuclear data libraries, such as those included in EASY-II(12) \cite{SUBLET1, TENDL11}, which is a functional replacement and enhancement to previous European Activation System releases.

\section{Experimental activities}

A selection of results from two recent experimental campaigns at the ASP facility are detailed in this paper. Example details are provided of integral cross sections derived from gamma spectrometry measurements of activated foils using a calibrated high-purity germanium (HPGe) detector. Elemental foils, typically of 12 mm diameter and 0.125 mm thickness, were irradiated with neutron fluxes at the sample irradiation position of approximately 10\textsuperscript{9} n cm\textsuperscript{-2} s\textsuperscript{-1} for irradiation times ranging from a few minutes up to about 1 hour, depending on the experiment. Following rapid extraction of the foils from the irradiation field via a `rabbit' transfer system, gamma spectra were recorded using a multi-channel analyser (MCA). The recorded spectra were typically sub-divided into 5 second time bins to enable identification of and/or separation of interfering gamma lines on the basis of half-life. Measurements from Al and Fe reference foils, which were stacked with the foil(s) of interest, were used to accurately determine the neutron fluence at the irradiation position for each experiment. In addition, two \textsuperscript{238}U fission counter readings provided a functional check of neutron fluence.

\subsection{Data acquisition and processing for W irradiation experiments}

Basic data acquisition for each experiment was managed using `job' files---a set of commands stored in a file that are sent to the MCA prior to measurement and used to separate gamma energy spectra into time bins. The large number of gamma spectrum files generated over the experimental campaigns meant that processing by-hand was impractical. In this work it was necessary to also develop automatic post-processing schemes that would allow the complete set of results to be analyzed in an efficient manner. For example, the scheme calculates decay-corrected activities based on the entire measurement history needed for cross section calculations, which are preferable to activities based on a time integrated single peak measurement.

\begin{figure}[htbp]
 \centering
 \includegraphics[width=1.0\linewidth]{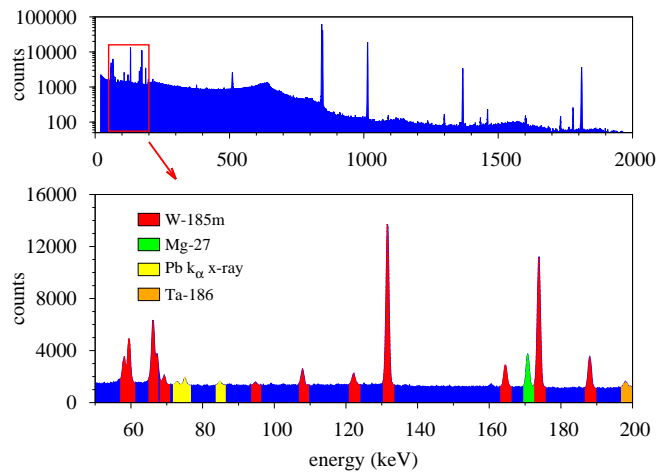}
 \caption{Gamma energy spectrum, in keV, measured shortly after a W foil neutron irradiation.}
 \label{fig:W_plot_eps}
\end{figure}

Here, an illustration is made by presenting data that was recorded for one of the W irradiation experiments. Figure \ref{fig:W_plot_eps} shows a time integrated gamma energy spectrum for a W irradiation, over both the entire 2 MeV energy range, and, at higher image resolution, the low energy portion of the spectrum up to 200 keV. One can clearly see the peaks that are associated with the \textsuperscript{185m}W reaction product. Also evident in the low energy region of the spectrum is a gamma line at 170.4 keV from the \textsuperscript{27}Mg product due to activation of the Al reference foil, that was used (along with a Fe foil) to determine the experimental neutron fluence, and some Pb x-rays originating from the detector shielding.  

\begin{figure}[htbp]
 \centering
 \includegraphics[width=1.0\linewidth]{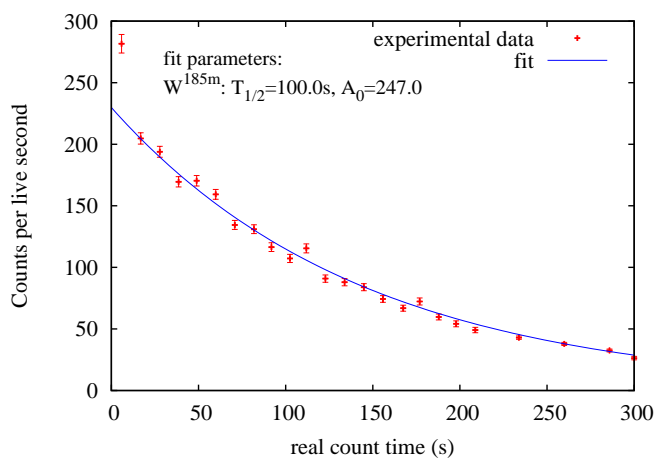}
 \caption{Activity as a function of time for the \textsuperscript{185m}W reaction product, derived from a gamma line measurement at 132 keV measured shortly after neutron irradiation of a W foil. The A0 value shown gives the decay corrected activity to the end of the irradiation period, which has been derived from the fitted line. A half-life of 100 s, taken from the EAF-2007 library, was used in the exponential fit.} 
 \label{fig:W-185mdecay132}
\end{figure}

A post-processing code was developed to read and manipulate the data contained within each spectrum file. This code was extended and modified so that it could analyze the sequence of spectra files produced by the job files for a given experiment, and track the evolution of a particular peak of interest. For a specified peak and sequence of spectra files, the code takes, as a user-input, various parameters, including: the number of channels in each spectrum; the initial and final measurement times to analyze, as well as the interval between each spectrum; and the peak energy and intensity for the gamma line of interest taken from the EAF decay library.
For the W experiment shown in Figure \ref{fig:W_plot_eps} it is possible to track the decay of the \textsuperscript{185m}W reaction product for the 132 keV line over the duration of the measurement. This is shown in Figure \ref{fig:W-185mdecay132}. The EAF-2007 decay data half-life value of 100 s for \textsuperscript{185m}W gives a good fit to the majority of the data. The outlying data point shown for the first time bin was excluded from the fitted line. An outlying data point was observed for the first time bin in several other experiments as well. Whilst the data could be a real effect, it is perhaps more likely that some small PC--MCA communication timing issues at the start of the aquisition period exist, but this requires more investigation. Either way, this illustrates the benefits of the time-bin approach over the use of time-integrated measurements---in the latter case the anomalous data at the start of counting would have un-knowingly been included.

The code stores and uses recorded experimental parameters and, for each gamma spectrum, automatically calculates peak widths and corrects for the Compton-background component of the spectrum using standard methods. Activity is calculated using the HPGe detector efficiency function and a user-defined intensity of the corresponding gamma-line (taken from EAF libraries, for example). Integral cross sections for reactions are calculated from the activity using standard activation techniques. The neutron flux was derived using EAF-2010 cross sections associated with standard reference foils, Al and Fe, and the calculated facility neutron spectrum at the irradiation position \cite{Packer2012}. In the case of experiments exhibiting multiple pathways to an identical reaction product, reactions are currently noted as summed reactions. Development plans for the code include separating summed reactions into individual component reactions using FISPACT pathway analysis.

\begin{figure}[htbp]
 \centering
 \includegraphics[width=0.95\linewidth,  angle=0]{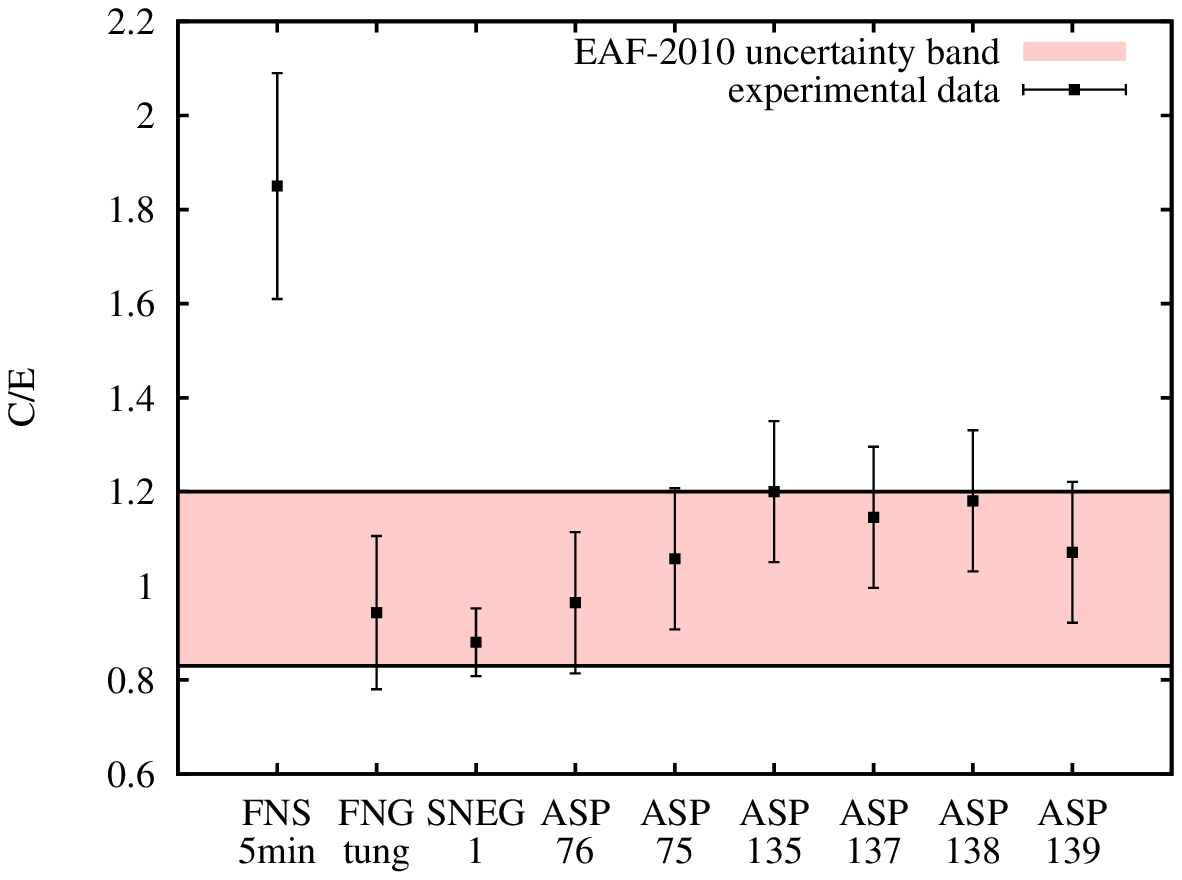}
 \caption{C/E plot for the \textsuperscript{186}W(n,2n)\textsuperscript{185m}W reaction.}
 \label{fig:CoverEW}
\end{figure}

Figure \ref{fig:CoverEW} shows a C/E plot for the \textsuperscript{186}W(n,2n)\textsuperscript{185m}W reaction. C/E plots have been used as part of the evaluation and validation processes used to produce the EAF data libraries. The activity of a measured nuclide is compared with a calculated value using the library data. The reaction producing the nuclide is identified and its average cross-section in the neutron spectrum is used as E. Similarly the library data are averaged in the neutron spectrum to form C. The experimental uncertainty has been estimated from the HPGe measurement statistics, detector calibration and neutron spectrum uncertainties; these are shown by error bars. The EAF-2010 library uncertainty for this reaction is shown by the error band. 

Figure \ref{fig:CoverEW} gives a C/E comparison of data derived from six separate measurements taken at ASP. These data points have error bars between 13-15 \% and are compared with those derived from experiments at the Frascati Neutron Generator (FNG) facility \cite{FNG}, the Neutron Source (FNS) facility in Japan \cite{FNS} and the SNEG facility in St Petersburg, Russia \cite{SNEG}. In the plot one can see that most of the data points lie within the EAF-2010 evaluation uncertainty band. The ASP data for this reaction suggests good repeatability over the experiments, which were taken from two experimental campaigns separated by several weeks.

C/E plots are used in conjunction with differential data in the evaluation process which result in improvements in subsequent library releases. Figure \ref{fig:W-186(n,2n)xsdiff_col} shows differential cross section data for the \textsuperscript{186}W(n,2n)\textsuperscript{185m}W reaction. The curve shows the evaluation used in the EAF-2010 library, which is compared to differential measurements taken at different neutron energies. The broken curves give the uncertainty associated with the evaluated curve.

\begin{figure}[htbp]
 \centering
 \includegraphics[width=0.7\linewidth, angle=-90]{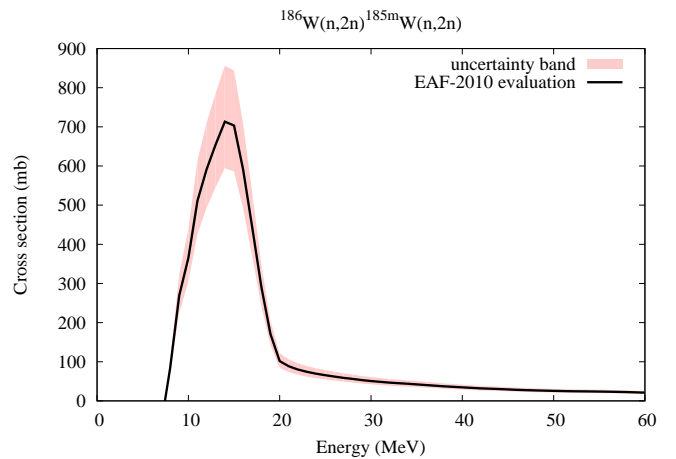}
 \caption{Differential cross section of the \textsuperscript{186}W(n,2n)\textsuperscript{185m}W reaction (from EAF-2010) and associated uncertainty band.}
 \label{fig:W-186(n,2n)xsdiff_col}
\end{figure}

\subsection{C/E values for range of experiments}

\begin{figure}[htbp]
 \centering
 \includegraphics[width=1.0\linewidth]{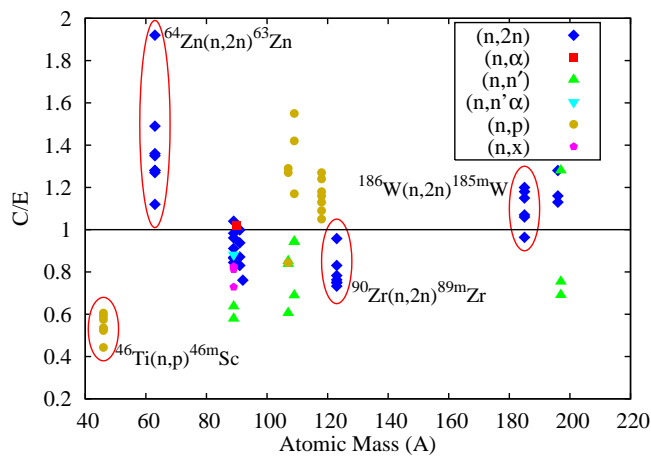}
 \caption{C/E values plotted against the atomic mass of the measured reaction product for a range of experiments. Values have been distinguished by reaction type.}
 \label{fig:CoverE_Aplot}
\end{figure}

Whilst it is not possible to present all measurements in detail here, a range of C/E values are shown in figure \ref{fig:CoverE_Aplot}. These are plotted against the atomic mass of the reaction product to show the range of measurements taken in the two campaigns. The C/E values have been distinguished by reaction type. In total 73 C/E values from 18 separate reaction types from Zn, Au, W, Zr, Sn, Ag, Nb, Y, Ti and Mo target foils are shown. Some groups of measurements are highlighted in the figure: (1) the \textsuperscript{186}W(n,2n)\textsuperscript{185m}W reaction, which were discussed earlier in the paper; (2) \textsuperscript{90}Zr(n,2n)\textsuperscript{89m}Zr; and (3) measurement for the \textsuperscript{46}Ti(n,p)\textsuperscript{46m}Sc reactions, for which there were no previous integral cross section measurements documented in \cite{EAFVAL}. The group of measurements for \textsuperscript{64}Zn(n,2n)\textsuperscript{63}Zn reactions display a fairly large scatter of C/E values. The relatively long half-life of \textsuperscript{63}Zn (38.4 minutes), compared with the measurement time, combined with the fairly weak gamma emission probability at 669.7 keV (0.0848), suggest that these measurements would have benefited from a longer spectrum acquisition period, a larger mass target foil and/or a higher fluence irradiation. It should be noted that the focus of these experiments was to measure short half-life products and that further optimisation of experimental method is necessary for future experimental campaigns to measure longer lived products.

Ongoing work to further improve the estimate of the ASP facility spectrum using neutron spectrometry and unfolding techniques is to be reported in conjunction with this work and may help to further improve C/E assessments in the future.

The availability of a single HPGe detector at the ASP facility meant that each irradiation--measurement experiment has been conducted as a serial process. This combined with the aim to cover a large number of short experiments over the campaign presented some practical timing constraints. For this reason it has been necessary to focus on short-lived reaction products in this work with the expectation that longer-lived measurements will be conducted over a longer period at a later date. Most of the reactions shown in figure \ref{fig:CoverE_Aplot} were derived from the measurement of short-lived reaction products, ranging from the 7.47 s half-life associated with the \textsuperscript{197m}Au reaction product to the 10.15 day half-life associated with the \textsuperscript{92m}Nb reaction product. Several of the long-lived reaction products of interest were evident in the measured spectra, but many were deemed to have not been measured with adequate counting statistics, so are not presented here (as discussed earlier, only the\textsuperscript{64}Zn(n,2n)\textsuperscript{63m}Zn measurements are presented to highlight this). It is expected that the measurement of gamma emissions from some longer-lived reaction products produced in activated foils will be measured at a later date at the CCFE radiological metrology lab.

\section{Conclusions and future work}

This work presents new integral data measured recently at the ASP 14 MeV neutron irradiation facility at Aldermaston in the UK. Such experiments are needed to contribute to the pool of experimental data to support activation data libraries, including those used in EASY-II(12), a functional replacement and enhancement to previous European Activation System releases.

A selection of results from the two most recent experimental campaigns at ASP are presented, which include details of the experimental method, example measurements and status of post-processing tools that have been developed. The approach is illustrated for the \textsuperscript{186}W(n,2n)\textsuperscript{185m}W reaction. 

A range of C/E values were derived from the measurements that were taken, using the post-processing tools that were developed. These values are presented and include new integral measurements for the \textsuperscript{46}Ti(n,p)\textsuperscript{46m}Sc reaction.

The measurements of some longer-lived reaction products from these experiments are planned to be conducted at the CCFE radiological metrology lab, which is expected to be operational in the near future.

\section{Acknowledgements}

This work was funded by the RCUK Energy Programme under grant EP/I501045 and the European Communities under the contract of Association between EURATOM and CCFE. The views and opinions expressed herein do not necessarily reflect those of the European Commission.
\bibliography{all}
\end{document}